%% file: main.tex
\documentclass[a4paper,11pt]{article}
\pdfoutput=1 
\usepackage{color}
\usepackage{jcappub} 

\usepackage{graphicx}
\usepackage{bm}
\usepackage{epsf}
\usepackage{rotating}
\usepackage{epsfig,graphics,rotate,color}
\usepackage{wrapfig}
\usepackage{amssymb}
\usepackage{amsmath}
\usepackage{amsfonts}
\usepackage{array,hhline,dcolumn}%
\usepackage{gensymb}
\usepackage{placeins}
\usepackage{tabularx}
\usepackage[colorlinks=true,citecolor=blue,linkcolor=blue]{hyperref}
\usepackage[capitalise]{cleveref}
\usepackage{comment}

\usepackage{siunitx}
\include{units}

\newcommand{\like}{\mathcal{L}}
\newcommand{\likeSAY}{\mathcal{L}_{\rm Eff}}

\author[16]{R. Abbasi,}
\author[60]{M. Ackermann,}
\author[17]{J. Adams,}
\author[11]{J. A. Aguilar,}
\author[21]{M. Ahlers,}
\author[50]{M. Ahrens,}
\author[22]{J.M. Alameddine,}
\author[30]{A. A. Alves Jr.,}
\author[42]{N. M. Amin,}
\author[40]{K. Andeen,}
% ...
\affiliation[11]{Universit{\'e} Libre de Bruxelles, Science Faculty CP230, B-1050 Brussels, Belgium}
% ...
\affiliation[16]{Department of Physics, Loyola University Chicago, Chicago, IL 60660, USA}
\affiliation[17]{Dept. of Physics and Astronomy, University of Canterbury, Private Bag 4800, Christchurch, New Zealand}
% ...
\affiliation[21]{Niels Bohr Institute, University of Copenhagen, DK-2100 Copenhagen, Denmark}
\affiliation[22]{Dept. of Physics, TU Dortmund University, D-44221 Dortmund, Germany}
% ...
\affiliation[30]{Karlsruhe Institute of Technology, Institute for Astroparticle Physics, D-76021 Karlsruhe, Germany }
% ...
\affiliation[40]{Department of Physics, Marquette University, Milwaukee, WI, 53201, USA}
% ...
\affiliation[42]{Bartol Research Institute and Dept. of Physics and Astronomy, University of Delaware, Newark, DE 19716, USA}
% ...
\affiliation[50]{Oskar Klein Centre and Dept. of Physics, Stockholm University, SE-10691 Stockholm, Sweden}
% ...
\affiliation[60]{DESY, D-15738 Zeuthen, Germany}
% ...

\date{\today}

\collaboration{IceCube Collaboration}

\title{Searches for Connections between Dark Matter and High-Energy Neutrinos with IceCube}

\abstract{
In this work, we present the results of searches for signatures of dark matter decay or annihilation into Standard Model particles, and secret neutrino interactions with dark matter.
Neutrinos could be produced in the decay or annihilation of galactic or extragalactic dark matter.
Additionally, if an interaction between dark matter and neutrinos exists then dark matter will interact with extragalactic neutrinos.
In particular galactic dark matter will induce an anisotropy in the neutrino sky if this interaction is present.
We use seven and a half years of the High-Energy Starting Event (HESE) sample data, which measures neutrinos in the energy range of approximately $\SI{60}\TeV$ to $\SI{10}\PeV$, to study these phenomena.
This all-sky event selection is dominated by extragalactic neutrinos. 
For dark matter of $\sim\SI{1}\PeV$ in mass, we constrain the velocity-averaged annihilation cross section to be smaller than $10^{-23}{\si\cm^3/\si\s}$ for the exclusive $\mu^+\mu^-$ channel and $10^{-22}{\si\cm^3/\si\s}$ for the $b\bar b$ channel.
For the same mass, we constrain the lifetime of dark matter to be larger than $10^{28}{\si\s}$ for all channels studied, except for decaying exclusively to $b\bar b$ where it is bounded to be larger than $10^{27}{\si\s}$.
Finally, we also search for evidence of astrophysical neutrinos scattering on galactic dark matter in two scenarios.
For fermionic dark matter with a vector mediator, we constrain the dimensionless coupling associated with this interaction to be less than 0.1 for dark matter mass of $\SI{0.1}\GeV$ and a mediator mass of $10^{-4}~{\si\GeV}$.
In the case of scalar dark matter with a fermionic mediator, we constrain the coupling to be less than 0.1 for dark matter and mediator masses below $\SI{1}\MeV$.

}

\begin{document}

\maketitle

\section{Introduction\label{sec:intro}}

Despite numerous dedicated experimental searches, the elusive nature of dark matter~\cite{Bertone:2004pz} remains one of the most important problems in modern physics.
High-energy neutrinos provide a new gateway to search indirectly for dark matter.
The recent observation of a high-energy astrophysical neutrino flux by the IceCube Neutrino Observatory~\cite{Aartsen:2013bka,Aartsen:2013jdh,Aartsen:2014gkd,Aartsen:2015rwa,Aartsen:2016xlq,Ahlers:2018fkn,Abbasi_2021}, whose origin remains to be understood, brings a unique opportunity to search for new phenomena~\cite{Arguelles:2019rbn}.
The IceCube data sample allows us to test many previously experimentally inaccessible dark matter models and probe relations between the dark matter and the neutrino sector~\cite{sergio_palomares_ruiz_2020_3959654}.

Studies of processes where self-annihilating dark matter captured in the Sun or Earth~\cite{Kamionkowski:1991nj, Bottino:1991dy, Halzen:1991kh, Gandhi:1993ce, Bottino:1994xp, Bergstrom:1996kp, Bergstrom:1998xh, Barger:2001ur, Bertin:2002ky, Hooper:2002gs, Bueno:2004dv, Cirelli:2005gh, Halzen:2005ar, Mena:2007ty, Lehnert:2007fv, Barger:2007xf, Barger:2007hj, Blennow:2007tw, Liu:2008kz, Hooper:2008cf, Wikstrom:2009kw, Nussinov:2009ft, Menon:2009qj, Buckley:2009kw, Zentner:2009is, Ellis:2009ka, Esmaili:2009ks, Ellis:2011af, Bell:2011sn, Kappl:2011kz, Agarwalla:2011yy, Chen:2011vda, Kundu:2011ek, Rott:2011fh, Das:2011yr, Kumar:2012uh, Bell:2012dk, Silverwood:2012tp, Blennow:2013pya, Arina:2013jya, Liang:2013dsa, Ibarra:2013eba, Albuquerque:2013xna, Baratella:2013fya, Guo:2013ypa, Ibarra:2014vya, Chen:2014oaa, Blumenthal:2014cwa, Catena:2015iea, Chen:2015uha, Belanger:2015hra, Heisig:2015ira, Danninger:2014xza, Blennow:2015hzp, Murase:2016nwx, Lopes:2016ezf, Baum:2016oow, Allahverdi:2016fvl, Rott:2012qb, Bernal:2012qh, Rott:2015nma, Rott:2016mzs,Aartsen:2016zhm,Choi:2015ara,Adrian-Martinez:2016gti,Albert:2016dsy,Aartsen:2016fep} have resulted in competitive bounds on the dark matter-nucleon scattering cross section.
Searches for self-annihilating dark matter from the galactic halo near the galactic center have constrained the dark matter self-annihilation cross section~\cite{Abbasi:2011eq,Arg_elles_2021,Albert_2020}.
Despite the progress in searching for high-energy and low-energy neutrinos of \textit{scotogenic} origin~\cite{Choi:2019zxy,Abada:2014zra,Ballett:2019cqp,Bhattacharya:2018ljs,Chianese:2018dsz,de_Boer_2021} there is so far no compelling signal for dark matter in neutrinos~\cite{Bertone:2004pz,Undagoitia:2015gya,Arcadi:2017kky,Arg_elles_2021}.
In this work, we perform a new analysis on the highest energy neutrinos observed so far, looking for evidence of a connection between dark matter and high-energy neutrinos.

The IceCube Neutrino Observatory~\cite{Aartsen:2016nxy} which is located at the geographic South Pole consists of 5160 digital optical sensor modules (DOMs)~\cite{Abbasi:2008aa,Abbasi_2010} distributed over 86 strings installed between $\SI{1.5}\km$ and $\SI{2.5}\km$ below the surface of the Antarctic ice~\cite{Aartsen:2013rt}.
A surface air-shower array, called IceTop~\cite{2013NIMPA.700..188A}, complements the in-ice detector by performing dedicated measurements of the cosmic-ray spectrum and providing rejection power for cosmic-ray events.
In the ice, vertical strings of DOMs are separated by about $\SI{125}\meter$ horizontally and on each string DOMs are spaced $\SI{17}\meter$ apart vertically.
A denser sub-array located in the bottom center of the detector is referred to as DeepCore~\cite{Collaboration:2011ym} and gives access to neutrino events below $\SI{100}\GeV$.
This sub-array is important for low-energy neutrino measurements~\cite{Aartsen:2017nmd}, but is not of primary interest in this paper.

In this work, we report results based on an IceCube data sample of high-energy starting neutrino events from a data-taking period of approximately $\SI{7.5}\year$.
This data sample has been analyzed using improved modeling of the detector response and the ice properties~\cite{Abbasi_2021}.
We investigate two dark matter scenarios: one where decaying or self-annihilating dark matter with masses above $\SI{100}\TeV$ produced neutrinos, and a second one with dark matter-neutrino interactions for dark matter masses below $\SI{1}\GeV$.

In the first analysis, we search for signatures of dark matter self-annihilation or decay, that directly or indirectly produce high-energy neutrinos, in the galactic center, halo, and with extragalactic origins~\cite{hrvoje_thesis}.
This signal could emerge as a component of the astrophysical neutrino flux with a specific directional distribution and energy spectrum~\cite{Beacom:2006tt,Yuksel:2007ac,PalomaresRuiz:2007ry}.
The directional distribution is defined by the galactic dark matter halo profile~\cite{Iocco:2015xga,2019JCAP...09..046K,Salucci_2019}, while the energy spectrum depends on the dark matter mass and annihilation channels; see~\cite{Jungman:1995df} for an extended discussion.
The analysis presented uses both the directional and energy distribution, which results in a more powerful --- though more model-dependent --- search than only using the directional information~\cite{Basegmez_du_Pree_2021}.

In the second analysis, we search for signatures of dark matter-neutrino interactions that occur as astrophysical neutrinos travel from their extragalactic sources to the Earth.
Along the way they can interact with extragalactic dark matter producing a distinct distortion in the energy spectrum~\cite{Arguelles:2017atb} and potentially altering the flavor composition~\cite{Farzan:2018pnk,Miranda_2015,de_Salas_2016}.
As they get closer to us, they interact with the galactic dark matter halo.
This interaction results in high-energy neutrinos losing energy, which creates an anisotropy in the neutrino sky and attenuates the high-energy neutrino flux~\cite{Arguelles:2017atb}, which is assumed to be originally isotropic due to the random distribution of the far away sources in the sky.
In this work, we only consider anisotropy produced on the diffuse flux as it is less model-dependent on the astrophysical source spectrum and is agnostic of the source distance.
For a discussion of the signatures on individual sources see~\cite{Kelly:2018tyg}.
All analyses described in this work were performed both in frequentist and Bayesian statistical frameworks.


Neither of the two analyses outlined above resulted in any evidence for a signal due to dark matter. 
Our results significantly improve upon existing experimental limits on the dark matter self-annihilation cross-sections, dark matter lifetimes, and dark matter-neutrino scattering cross-sections.
The rest of this article is structured as follows: in Sec.~\ref{sec:event_selection} we describe the event selection used; in Sec.~\ref{sec:signal} we define the signatures of dark matter annihilation, decay, and neutrino scattering; in Sec.~\ref{sec:analysis} we describe the analysis strategy and in Sec.~\ref{sec:results} we present the results; and in Sec.~\ref{sec:conclusion} we conclude.


\section{Event selection\label{sec:event_selection}}

This analysis uses the event selection first introduced in~\cite{Aartsen:2013jdh,Aartsen:2014gkd} and recently updated in~\cite{Abbasi_2021} with an updated model of the Antartic ice, a re-calibrated detector response to single photo-electrons~\cite{IceCube:2020nwx}, and an improved treatment of systematic uncertainties associated with the neutrino flux and detector response.
We briefly review the event selection, highlighting the points most relevant for the analyses performed in this work.
Further details on the event selection are provided in~\cite{Abbasi_2021}.
The high-energy starting event (HESE) selection uses the outer part of the detector as a veto to reduce the atmospheric neutrinos and muon backgrounds.
The veto is formed from signals in the `veto-DOMs' which are all DOMs in the outermost layer of strings and, for most strings, the top six layers of DOMs.
Included in the veto is also the bottom layer of DOMs.
A few strings were installed a little lower or higher than their neighbors and the selection of veto DOMs at the top and bottom is adapted accordingly.
In addition, the DOMs in a $\SI{60}\meter$ thick horizontal layer located approximately $\SI{2100}\meter$ deep are included in the veto as this region has a higher concentration of dust.
The HESE selection requires events to satisfy two criteria:
\begin{itemize}
    \item no more than three photo-electrons (PE) in the DOMs in veto region causally connected with the reconstructed vertex,
    \item and a minimum of $\SI{6000}\pe$ charge deposited in the detector volume.
\end{itemize} 
More precise definitions of these criteria are given in~\cite{Abbasi_2021}.
This combination of selection criteria selects events with a neutrino interaction vertex within the fiducial volume.
The effective volume of this event selection is approximately independent of the incoming neutrino direction resulting in an effective target mass on the order of $200~\si\Mtons$~\cite{Palomares-Ruiz:2015mka,Abbasi_2021} above $\SI{100}\TeV$.

The angular distribution of the expected extraterrestrial neutrino induced rate is dependent on the angular distribution of the flux incident on Earth and the Earth's opacity to high-energy neutrinos.
The flux incident on the Earth depends on the assumed astrophysical neutrino origin~\cite{10.3389/fspas.2019.00032}.
In this work, we consider two types of incident astrophysical neutrino fluxes: one that is assumed isotropic, and another that is anisotropic and originates from dark matter clustered around the galactic center.
The Earth opacity depends on the high-energy neutrino cross-section~\cite{Gandhi:1995tf,CooperSarkar:2011pa} and the properties of the Earth~\cite{Dziewonski:1981xy,Donini:2018tsg}.
This opacity results in a decreased yield of neutrinos from directions traversing the Earth~\cite{Gandhi:1995tf}.
The uncertainty in the opacity is small given that at the energies relevant for this analysis, the cross-section is well-understood and the impact of the Earth density uncertainties in the opacity are small~\cite{Vincent:2017svp,Aartsen:2017kpd,Aartsen:2020fwb}.

The angular acceptance of atmospheric neutrinos and muons produced in cosmic-ray showers in the Earth's atmosphere is modified from its production angular distribution by the Earth opacity and detector overburden, respectively.
Furthermore, in the case of atmospheric neutrinos, as noted in~\cite{Schonert:2008is}, muons produced alongside neutrinos in cosmic-ray showers can trigger the veto layer, modifying the expected angular acceptance of atmospheric neutrinos.
Atmospheric muons with an incident energy of $\SI{10}\TeV$ trigger the veto with a probability of approximately $\SI{90}\percent$, while muons at $\SI{100}\TeV$ trigger the veto approximately $\SI{99}\percent$ of the time.
The high-efficiency of muon rejection is related to a significant suppression of atmospheric neutrinos in the Southern sky, where the overburden is small enough that the muons reach the detector.
This suppression was parameterized for all neutrino flavors for the first time in~\cite{Gaisser:2014bja}, and a more recent discussion given in~\cite{Arguelles:2018awr} has quantified its uncertainties and yields new parameterizations which we use in this work.

The suppression of atmospheric neutrinos in the Southern sky makes this sample unique, giving it an all-sky coverage with high astrophysical neutrino purity.
This suppression is particularly relevant for this work as the expected dark matter signatures are more pronounced in the direction of the galactic center which is in the Southern sky.
As IceCube is located at the South Pole, the background from atmospheric muons and neutrinos would be very strong without the veto.

\section{Signal expectations\label{sec:signal}}

In this section we describe how to calculate the yield of neutrinos from decay or annihilation of dark matter particles into Standard Model particles and the effect on the flux of astrophysical neutrinos caused by their scattering on dark matter particles.

\subsection{Dark Matter Decay or Annihilation}

In this work, we only look at the arrival directions, morphology, and deposited energies of the events, \textit{i.e.} we are agnostic to the neutrino production distance.
Thus the relevant quantity to infer the yield of neutrinos from dark matter, or the effects on the astrophysical neutrino flux, is either the dark matter density --- for decay or scattering --- or the square of the density  --- for annihilation --- integrated along the line of sight (l.o.s.).
This is given by
\begin{equation}
\mathcal{J}^\alpha_\chi(\delta,RA)=\int\limits_{\rm l.o.s.}\left[\rho^{G}_\chi(s)\right]^\alpha~ds,
\label{eq:J_factor}
\end{equation}
where $\alpha$ can take values one or two depending on whether we consider decaying or annihilating dark matter; $s$ parameterizes the distance from Earth along the line of sight defined by the declination, $\delta$, and the right ascension, $RA$; and $\rho^{G}_\chi(s)$ is the dark matter galactic density.
The Milky Way dark matter halo density distribution has been inferred from measurements of star rotation curves~\cite{Iocco:2015xga,2019JCAP...09..046K}, which has greater uncertainties close to the galactic center where the dark matter density is largest.
For definiteness, we assume an Einasto halo profile with parameters and parametrization given in~\cite{Vincent_2012}.
Then the neutrino flux at Earth from galactic dark matter decay is given by 
\begin{equation}
\left(\frac{d\Phi}{dE_{\nu}}\right)_{dec.} = \frac{1}{4\pi m_\chi\tau_\chi} \frac{dN_{\nu}}{dE_{\nu}} \mathcal{J}^1_\chi(\delta,RA),
\label{eq:dec_gal_flux}
\end{equation}
where $\tau_\chi$ is the dark matter lifetime and $\frac{dN_{\nu}}{dE_{\nu}}$ is the neutrino production spectrum.  Similarly, the flux of neutrinos from galactic dark matter annihilation is given by
\begin{equation}
\left(\frac{d\Phi}{dE_{\nu}}\right)_{ann.} = \frac{\langle  \sigma v\rangle}{8\pi m_{\chi}^2}  \frac{dN_{\nu}}{dE_{\nu}}\mathcal{J}^2_\chi(\delta,RA),
\label{eq:ann_gal_flux}
\end{equation}
where $\langle  \sigma v\rangle$ is the velocity-averaged dark matter annihilation cross section.

For the dark matter annihilation the sensitivity of the analysis primarily originates from the morphology of the signal, which is dominated by the Galactic component. 
The isotropic extragalactic contribution originating from distant sources is neglected.
Given our data driven approach, we consider an isotropic extragalactic component as part of our background estimate, which would increase the overall background making the analysis less sensitive and therefore results in a less constraining limit. 

\begin{figure}[h]
    \centering
    \includegraphics[width=0.45\columnwidth]{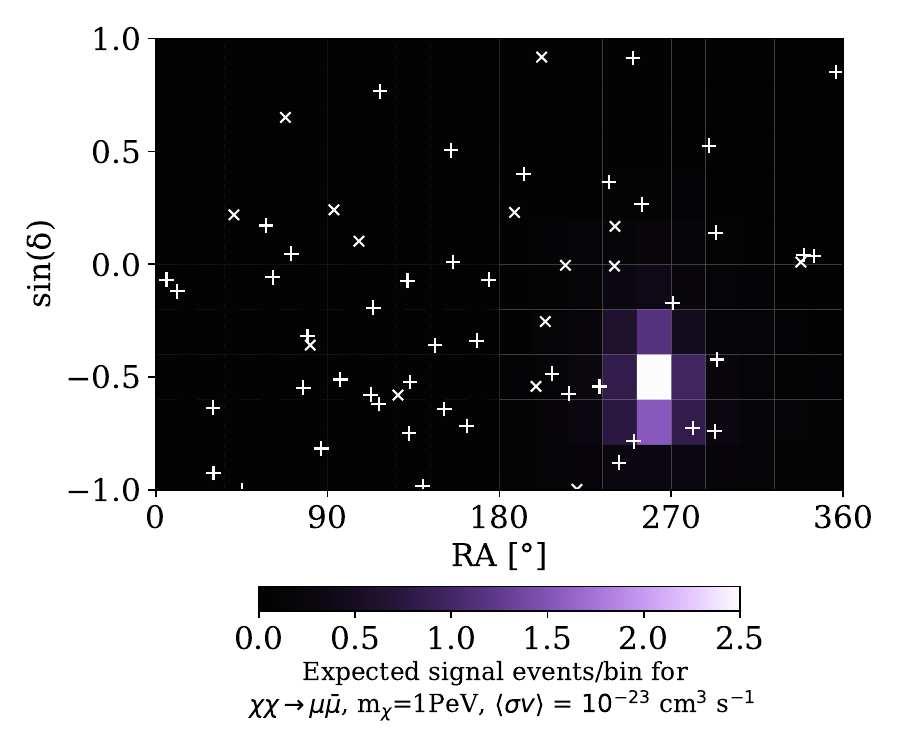}
    \includegraphics[width=0.45\columnwidth]{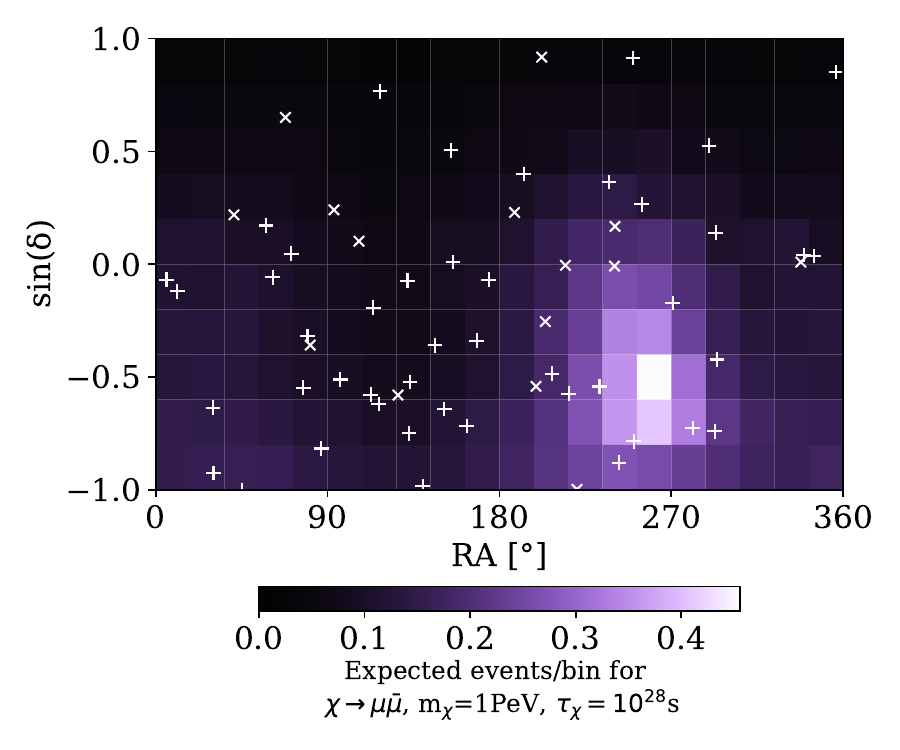}
	\protect\caption{\textbf{\textit{Expected distribution of events from dark matter annihilation and decay.}}
	The sky distribution of the expected signal contribution due to dark annihilation (left) and decay (right).
	For the dark matter annihilation and decay models, the expected events are shown per bin for the same binning as was used in the analysis.
	In this figure, the dark matter mass is assumed to be $\SI{1}\PeV$ and the annihilation or decay channel is set to be $\mu\bar{\mu}$, with the following parameters $\langle\sigma v \rangle = 10^{-23} {\rm cm}^3$ s$^{-1}$ and $\chi\rightarrow\mu\bar{\mu}$, $m_\chi=\SI{1}\PeV$, $\tau_\chi=10^{28}{\rm s}$, respectively.
	The assumed annihilation/decay channel and the dark matter mass have little influence on the shape of the expected event sky distribution.
	The velocity-averaged cross-section and lifetime only affect the overall normalization.
	In these figures, we assume the Einasto dark matter profile.
	Scatter points indicate the best-fit directions of events present in the HESE sample, (+) indicates events classified as cascades, while (x) marks events identified as tracks.}
    \label{fig:ann_dec_skymaps}
\end{figure}

For the dark matter decay scenario, we do consider the extragalactic contributions, as for this case the morphology of the signal is less dominated by the Galactic component and more spread out as seen in Figure~\ref{fig:ann_dec_skymaps}. 
This is due to the wider angular resolution for the cascade morphology, which implies that the angular patch on the sky of the signal is larger. 
This implies that larger atmospheric backgrounds enter the region compared to the galactic contribution which is narrower.
The extragalactic contribution originating from cosmological dark matter and from dark matter in sources such as galaxy clusters, is assumed to be isotropic. 


We consider a diffuse isotropic astrophysical neutrino flux as part of our background. We assume a single power law flux with normalization and spectral index fit to the observed data.
As a conservative approach, we only take into account the cosmological dark matter contribution by adding the following contribution to the galactic neutrino flux
\begin{equation}
\frac{d\Phi}{dE_{\nu}}^{cos.}_{dec.} = \frac{\Omega_{\chi} \rho_{\textrm{c}}}{4\pi m_\chi\tau_{\chi} H_0} \int \limits_0^{\infty} \frac{dN_{\nu}}{d(E_{\nu}(1+z))}\frac{dz}{\sqrt{\Omega_{\Lambda}+\Omega_{\textrm{m}}(1+z)^3}},
\label{eq:dec_exgal_flux}
\end{equation}
where $z$ is the redshift, and the $\Lambda$CDM model parameters --- critical density ($\rho_c$), Hubble parameter ($H_0$), cold dark matter density ($\Omega_{\chi}$), matter density ($\Omega_m$), and energy density ($\Omega_\Lambda$) --- are taken as the best-fit values in Table 3 from~\cite{2016A&A...594A..13P}.

In order to predict the spectrum of neutrinos from dark matter decay or annihilation, namely $\frac{dN_{\nu}}{dE_{\nu}}$ used in Eqs.~\eqref{eq:ann_gal_flux},~\eqref{eq:dec_gal_flux}, and~\eqref{eq:dec_exgal_flux}, we proceed in the following way.
The neutrino flux in the center-of-mass frame of the decaying dark matter or annihilating dark matter pair is computed.
For dark matter decay, the \texttt{PYTHIA}-8.1~\cite{Sjostrand:2014zea} package is used to generate a meta-stable spin-zero particle with a fixed mass and letting it directly decay into one of several designated channels; for details see~\cite{hrvoje_thesis}.
\texttt{PYTHIA} partially accounts for contributions from electroweak corrections~\cite{Christiansen:2014kba} by considering the emission of $W$ and $Z$ bosons off fermions, but neglects, \textit{e.g.}, electroweak triple gauge couplings; see~\cite{Cirelli:2010xx} for an alternative calculation that includes these effects.
The computations for galactic dark matter in \cite{Bauer_2021} improve upon those of \cite{Cirelli:2010xx} and apply in vacuum, or in general, in low-density media, whereas the adapted calculations for Earth and Solar dark matter neutrino fluxes in \cite{Liu:2020ckq} also account for the absorption of the annihilation products in high-density media.
These contributions are important at our energy scale and imply that the neutrino yield, specially of low-energy neutrinos, could be higher making our result conservative.

Final-state neutrinos from these processes are counted, and their average energy distributions are calculated for each flavor, and per neutrino and antineutrino. 
For dark matter annihilation, we proceed similarly and assume the two annihilating dark matter particles combine into a meta-stable spin-zero state which then decays into Standard Model particles.
The resulting spectrum is then equivalent to the decay spectrum of a dark matter particle with double the mass.

\begin{figure}[th]
    \centering
    \includegraphics[width=0.5\columnwidth]{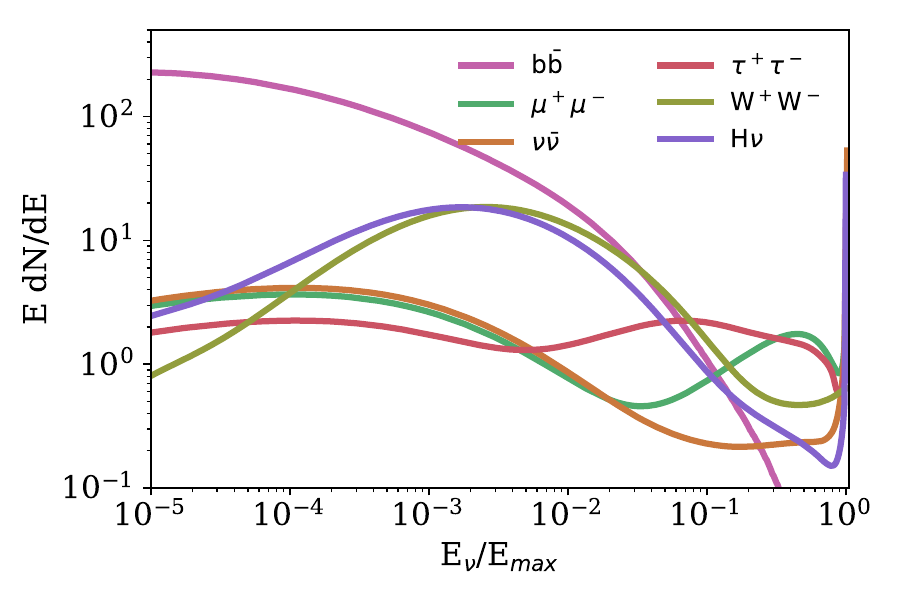}
	\protect\caption{\textbf{\textit{Neutrino distribution from dark matter decay or annihilation.}}
	Total neutrino and antineutrino per dark matter annihilation or decay for the different channels assuming a 100\% branching ratio to each channel.
	The maximum energy of the neutrino distribution is $E_\text{max}=m_\chi$ for the case of annihilation  and $E_\text{max}=m_\chi/2$ for decay, where the range of $m_\chi$ considered is $>10$ TeV.
	 The different colors specify different channels considered in this work. The spectra are shown at origin.
}
    \label{fig:energy_spectra}
\end{figure}

Since the dark matter mass in our energy range is much larger than all Standard Model particles produced in the decay or annihilation channels considered, the shape of the neutrino spectra is approximately independent of the dark matter mass, \emph{i.e.} the energy distribution $\frac{dN}{dx}$, where $x = E_\nu/E_\text{max}$, is similar in the case of soft coherence, at large $m_\chi$ and small $x$.
For a detailed discussion on the regimes of validity of this approximation see~\cite{Bauer_2021}.
This approximation is valid for dark matter masses in the range of $\SI{100}\TeV$ to $\SI{10}\PeV$ and has been checked by comparing the generated spectra for different dark matter particle masses.
It should be noted that this approximation breaks when additional electroweak processes are considered as discussed in~\cite{Bauer_2021}.
The resulting energy distribution, summed up over neutrino and antineutrino as well as all flavors, can be seen in Fig.~\ref{fig:energy_spectra}.
Given the energy of these neutrinos and the distances at which they originate, the resulting oscillation probability frequency is very large and many oscillations are contained within our detector energy resolution. 
This results in the oscillation probability to be averaged out and the transition probabilities are only given by mixing elements~\cite{Baerwald:2012kc,Mena:2014sja,Palomares-Ruiz:2015mka,Arguelles:2015dca,Bustamante:2015waa,Gonzalez-Garcia:2016gpq,Rasmussen:2017ert,Song_2021}.
Thus, the neutrinos are propagated to Earth assuming averaged oscillation, with best-fit oscillation parameters taken from~\cite{Esteban:2018azc}.
Examples of the expected distribution in the final analysis observables of this procedure can be seen in Fig.~\ref{fig:ann_dec_skymaps}.

\subsection{Dark Matter Scattering}
\begin{figure}[h]
    \centering
    \includegraphics[width=0.5\columnwidth]{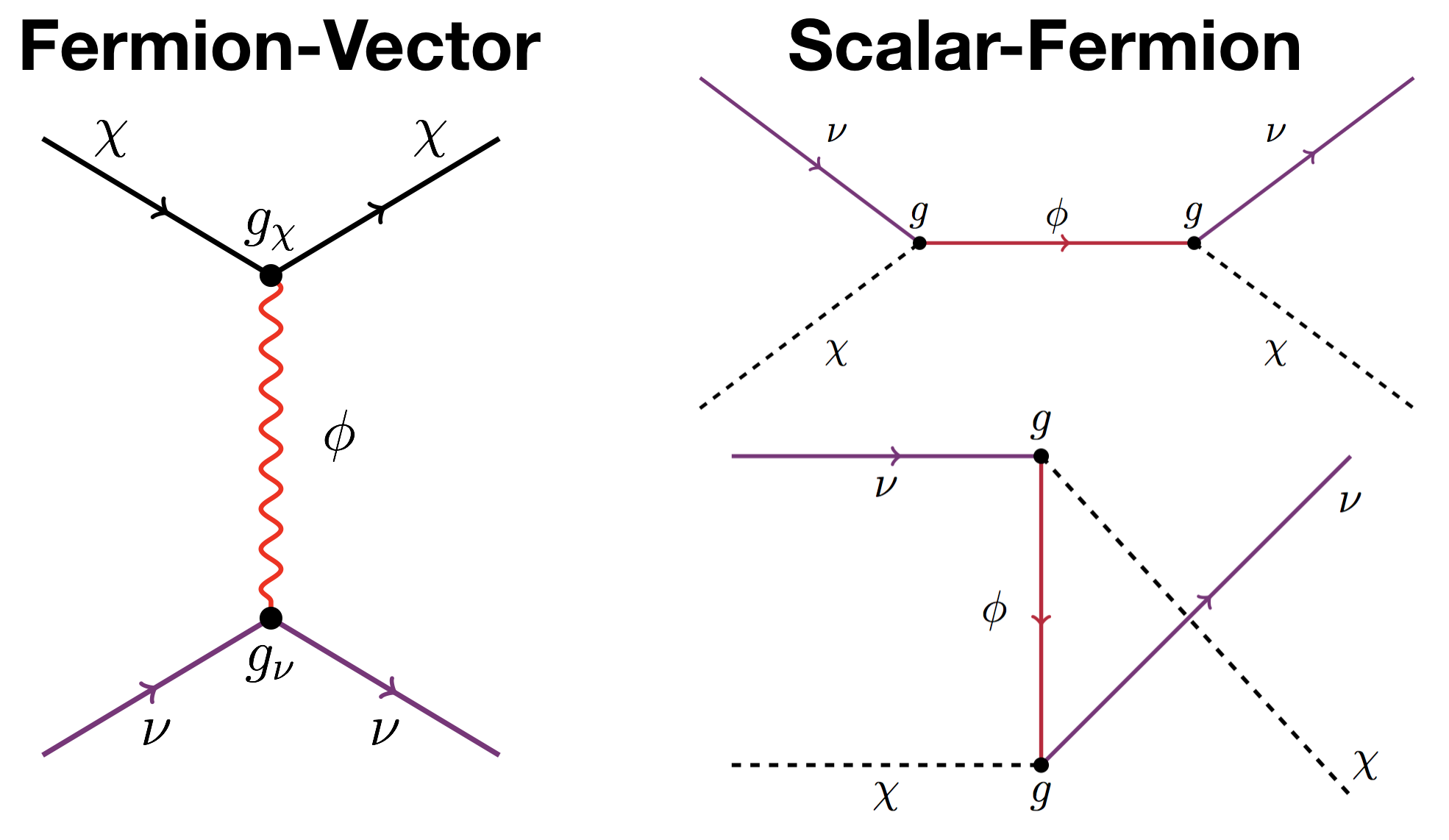}  
	\protect\caption{\textbf{\textit{Lowest order Feynman diagrams of dark matter-neutrino scattering for the processes considered.}} The left panel shows the fermionic dark matter with a vector mediator process and right one the scalar dark matter with a fermionic mediator.
	In these diagrams the dark matter ($\chi$) are shown in black, the mediator ($\phi$) in light red, and the neutrino ($\nu$) in purple.}
    \label{fig:feynman_scattering}
\end{figure}

For the dark matter-neutrino scattering analysis, we assume that dark matter, $\chi$, is either a scalar or fermion which couples only to a mediator particle, $\phi$.
Correspondingly, the mediator is assumed to be either a fermion or a vector and couples to neutrinos; see~\cite{Campo:2017nwh,Blennow_2019,Olivares-Del-Campo:2019qwe} for more details on these models. 
The leading order dark matter-neutrino scattering diagrams considered in this work are shown in Fig.~\ref{fig:feynman_scattering}.

In the case of fermionic dark matter with a vector mediator (FV) the dark matter-neutrino total cross-section is given by~\cite{Arguelles:2017atb,Olivares-Del-Campo:2019qwe}
\begin{eqnarray}
\sigma_{\chi\nu}^{FV}(E_\nu) = \frac{
g_\chi^2 g_\nu^2}{16\pi E_\nu^2 m_\chi^2}&&\left[ (m_\phi^2+m_\chi^2 + 2E_\nu m_\chi) \ln\left(\frac{m_\phi^2(2E_\nu+m_\chi)}{m_\chi(4E_ \nu^2+m_\phi^2)+2E_\nu m_\phi^2}\right)  \right. \\  &&+ \left. 4E_\nu^2 \left(1 + \frac{m_\chi^2}{m_\phi^2} - \frac{2E_\nu(4E_\nu^2m_\chi + E_\nu (m_\chi^2 + 2 m_\phi^2)+  m_\chi m_\phi^2}{(2E_\nu + m_\chi)(m_\chi(4E_\nu^2 + m_\phi^2 ) + 2E_\nu m_\phi^2)} \right) \right], \nonumber
\label{eq:xs_fv}
\end{eqnarray}
where $g_\chi$ is the $\chi\chi\phi$ vertex coupling strength and $g_\nu$ is the coupling strength for the $\phi\nu\nu$ vertex. 
For the scalar dark matter with a fermionic mediator (SF), the cross section is given by
\begin{eqnarray}
 \sigma_{\chi\nu}^{SF}(E_\nu) =  \frac{g^4}{64\pi} && \left[ \frac{8E_\nu^2 m_\chi}{(2E_\nu+m_\chi)(m_\phi^2-m_\chi(2E_\nu+m_\chi))^2} \right. \\
 &+&  \frac{4}{m_\chi^2 -2E_\nu m_\chi -m_\phi^2} +\frac{8}{m_\chi (2E_\nu + m_\chi )-m_\phi^2} \nonumber \\
&+&\left.  \left(\frac{4E_\nu m_\chi- 2(m_\chi^2+3m_\phi^2)}{E_\nu m_\chi(m_\chi(2E_\nu + m_\chi) - m_\phi^2)} + \frac{3}{E_\nu^2}\right)\ln \left(\frac{4E_\nu^2 m_\chi}{m_\phi^2(2E_\nu+m_\chi)-m_\chi^3} +1 \right) \right], \nonumber
\label{eq:xs_sf}
\end{eqnarray}
where $g$ is the coupling of the $\chi\phi\nu$ vertex.
The total interaction cross section together with the dark matter column density, defined in Eq.~\eqref{eq:J_factor} with $\alpha = 1$, determines the dark-matter-induced opacity, which is given by $\mathcal{S}(E_\nu, \delta, RA; m_\chi) = e^{-\sigma(E_\nu) \mathcal{J}^{1}(\delta, RA)/m_\chi}$.
Contribution from baryonic matter is neglected due to the small number density ($n_B =1 /{\rm cm}^3$) and the weak cross section involved. 
Thus, for a given dark matter mass, mediator mass, and couplings, the astrophysical flux is modified to be 
\begin{equation}
    \phi_{astro}(E_\nu, \delta, RA) = \mathcal{S}(E_\nu, \delta, RA) \phi_{astro}(E_\nu),
\end{equation}
where $\phi_{astro}(E_\nu)$ is assumed to be an isotropic flavor-democratic flux with an unbroken power-law energy distribution of extragalactic origin.
Note that in this approximation we neglect the secondary neutrinos from dark matter interaction, as these lower energy neutrinos give a subleading contribution given the soft astrophysical neutrino spectrum; see~\cite{Arguelles:2017atb} for a detailed discussion.
We also ignore the interaction with extragalactic dark matter due to the fact that this now depends on the point of origin of the neutrino and only produces an energy-spectrum signature; see~\cite{Kelly:2018tyg,Choi:2019ixb,Alvey_2019} for discussion on this topic and~\cite{Stecker:2017gdy} for a similar application in the context of Lorentz invariance violation.
The shadowing introduced by extragalactic dark matter introduces an energy cut-off independent of the direction, which is hard to distinguish from a cut-off of the source spectrum.
For this reason, we take a conservative approach and consider only the galactic dark matter shadow which creates a signature in energy and direction that cannot easily be mimicked by unaffected astrophysical neutrinos.

\begin{figure}[h]
    \centering
    \includegraphics[width=0.6\columnwidth]{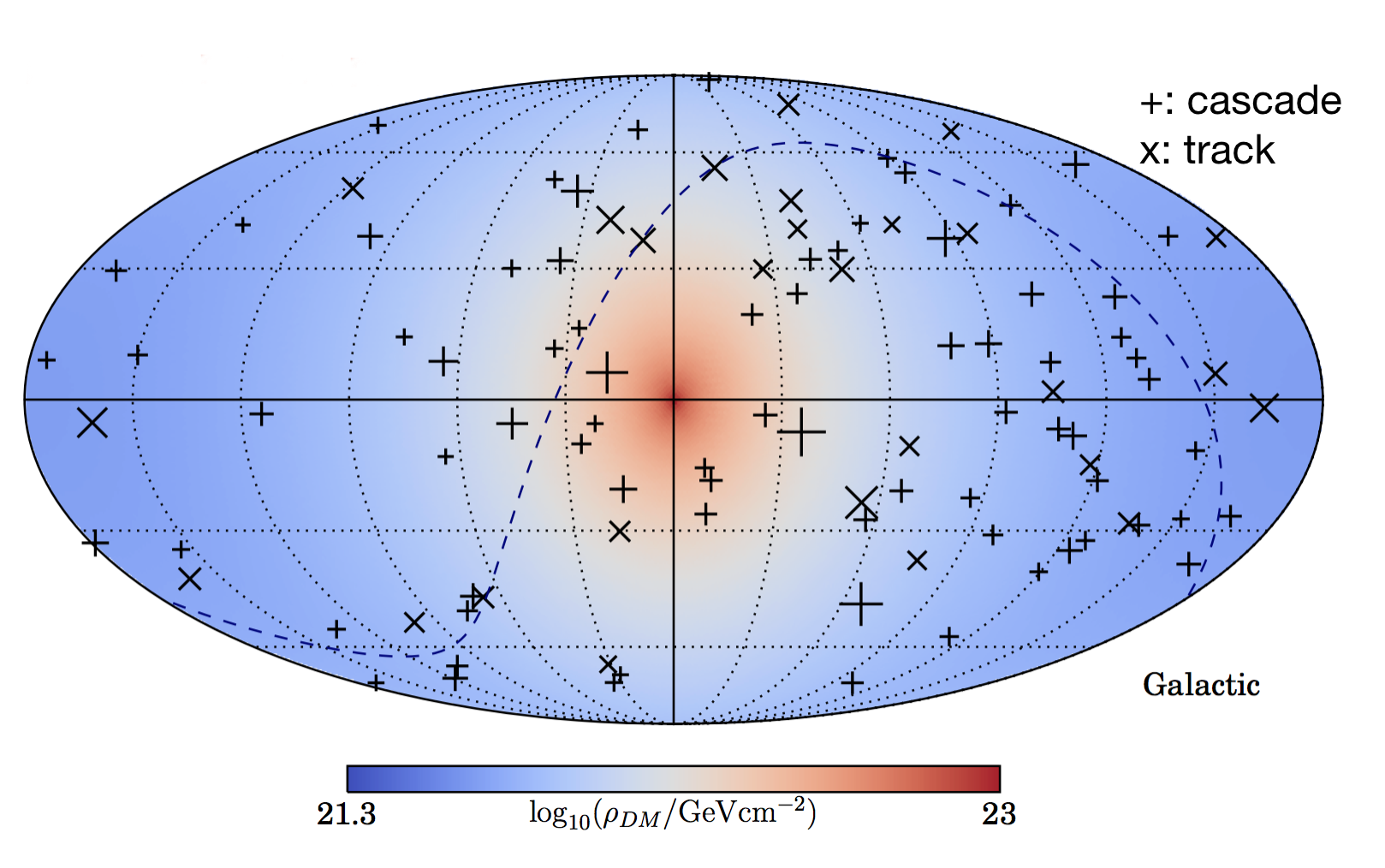}
	\protect\caption{\textbf{\textit{Distributions of events and the galactic dark matter density effect on dark matter-neutrino interactions.}} 
	In this figure, $\rho_\text{DM}$ denotes column density. The directions of the observed events, assuming all events are of extragalactic origin, are shown in the figures as (+) for (double-) cascade-like events and (x) for track-like events. Larger (smaller) markers indicate a relatively higher (lower) event energy. Figure is shown in galactic coordinates. Dashed line represents the horizon.}
    \label{fig:skymaps}
\end{figure}

Figure~\ref{fig:skymaps} shows the events best-fit locations superimposed with the dark matter column density.
The latter is correlated with the expected contributions to the overall neutrino signal from the dark matter annihilation, decay, and scattering models.

\section{Analysis method\label{sec:analysis}}

The observed events are assumed to be due to a combination of three components: muons and neutrinos produced in cosmic-ray air showers, neutrinos from conventional astrophysical sources, and neutrinos from dark matter.
The atmospheric component is divided into three pieces: the conventional atmospheric neutrino flux, which is dominated by kaon decay in the energy range relevant for this analysis; the prompt atmospheric neutrino flux dominated by $D$-meson decay; and the atmospheric muon flux.
The observed astrophysical neutrino flux is expected to be dominated by extragalactic contributions.
In fact, the galactic contribution has been constrained to be no more than $10\%$ of the observed astrophysical flux~\cite{Aartsen:2017ujz,Denton:2017csz,Albert:2018vxw}.
Its energy distribution is compatible with an unbroken single power-law and the flavor composition is expected to be equal for the preferred production mechanism~\cite{Palladino:2015vna,Arguelles:2015dca,Bustamante:2015waa}.
Thus, in this work, we assume the astrophysical component as an unbroken power-law spectrum and a democratic flavor composition, and equal amounts of neutrinos and antineutrinos.
The spectral parameters of the astrophysical neutrino flux are fitted for each signal hypotheses tested in this work.
For an in-depth discussion of the flux assumptions see~\cite{Abbasi_2021}.

Regarding the dark matter contribution to the astrophysical neutrinos, when considering dark matter decay or annihilation, the additional neutrino spectrum is modeled as discussed in Sec.~\ref{sec:signal}.
Finally, when considering the possibility of dark matter-neutrino scattering, we do not consider neutrinos produced from decays and self-annihilation of this dark matter as the dark matter masses that induce significant galactic opacity are much smaller than the neutrinos considered in this work.
Thus, in this latter case, we modify the energy and angular distributions of the astrophysical neutrino flux to account for this new force.

The data and Monte Carlo (MC) events are binned in morphology, $m$; reconstructed energy, E; declination, $\delta$; and right ascension, ${\rm RA}$.
We use twenty equal bins in $\log_{10}(E)$ between $\SI{60}\TeV$ and $\SI{10}\PeV$, ten bins in $\sin(\delta)$, twenty bins in ${\rm RA}$, and three bins in $m$ which correspond to tracks, cascades, and double cascades categories; see~\cite{Abbasi_2021} for more details regarding the bin choice.
The likelihood is a function of the dark matter model parameters, $\vec\Theta$, and nuisance parameters, $\vec\eta$, that take into account detector and flux uncertainties. 
The likelihood is defined as the product of the likelihood over all the bins, where the likelihood of the $i$-th bin when observing $x_i$ events with an expectation $\mu_i(\vec\Theta,\vec\eta)$ and associated MC sample size uncertainty $\sigma_i(\vec\Theta,\vec\eta)$, both derived from simulation, is given by the function $\mathcal{L}_{\rm Eff}$ defined in~\cite{Arg_elles_2019}. 
This likelihood function converges to a Poisson likelihood in the limit of infinite MC events and makes the likelihood shallower with increasing MC uncertainty. 
In symbols, the likelihood in the frequentist framework, or the unnormalized posterior density in the Bayesian framework, is given by
    \begin{equation}
        \like(\Theta,\eta) = \left[\prod_m\prod_i^{n_m} \likeSAY(\mu^m_i(\Theta,\eta), \sigma^m_i(\Theta,\eta); d^m_i)\right] \prod_{s} \Pi_s(\eta_s),
        \label{eq:likelihood}
    \end{equation}
where $\mu^m_i(\Theta,\eta)$ is the expected number of events in the morphology $m$-bin and $i$-th declination-RA-energy bin, $\sigma^m_i(\Theta,\eta)$ is the associated Monte Carlo uncertainty, $\Pi_s$ is the constraints (priors) on the nuisance parameters, and $n_m$ is the number of events per each of the three morphological categories, $m$.

In this work, we derive results using both the frequentist and Bayesian frameworks.
The Bayesian framework has the advantage that it treats systematic uncertainties and model parameters in the same way and reports credible upper limits on the model parameters. 
These are interpreted as the regions of parameter space that contain the highest posterior density for a given choice of model priors.
The frequentist framework treats systematic uncertainties differently than model parameters, as discussed below.
This method has the advantage that the confidence intervals constructed satisfy the coverage property.
Namely that upon repetition of the measurement the true value should be contained a certain fraction of the time in the interval, enabling comparison between different experiments.

For our frequentist results, we incorporate systematic uncertainties by constructing the profile likelihood, which is done by finding the nuisance parameters for each model parameter point that maximizes the likelihood function when its multiplied by terms that represent external constraints and information as defined in Eq.~\eqref{eq:likelihood}; see~\cite{Algeri_2020} for a recent discussion on this procedure.
Using the profile likelihood, we construct confidence intervals assuming Wilks' theorem~\cite{wilks1938}.
For our Bayesian results, we compute the posterior distribution of the physics and nuisance parameters using the EMCEE~\cite{ForemanMackey:2012ig} Markov-chain Monte Carlo sampler.
Then, we report the posterior distributions of our physics parameters marginalized over the nuisance parameters.
For more details and discussion on our statistical treatment see~\cite{Abbasi_2021}.

The sources of systematic uncertainties considered in this work, and implemented by means of nuisance parameters, encompass two categories: background modeling and detector response.
The flux uncertainties nuisance parameters allow for the normalization of each of the components to float with constraints.
These four independent normalizations correspond to the astrophysical flux, conventional atmospheric flux, prompt atmospheric neutrino flux, and atmospheric muon flux components.
The conventional atmospheric component is allowed to float with Gaussian constraint (prior) that represents the theoretical uncertainty in this energy range, the atmospheric muon component uncertainty is constrained by using a data-driven method, and finally the prompt atmospheric component and the astrophysical neutrino components are allowed to float freely, while keeping the normalizations positive as discussed in~\cite{Abbasi_2021}.
We take into account uncertainties in the cosmic-ray spectral index by allowing it to float, rescaling the atmospheric neutrino prompt and conventional components.
Additionally, we also vary the relative contribution between pions and kaons for the conventional component.

The detector uncertainties considered in this work are the absolute photon detection efficiency of the detector, various ice scattering and absorption models, and the direction and strength of the ice anisotropy.
These are incorporated as continuous parameterizations which linearly interpolate between dedicated Monte Carlo sets, which were constructed by varying one systematic uncertainty parameter at a time.
The fit is then allowed to change these detector parameters within priors motivated by \textit{in-situ} measurements.
These are found to be much smaller than the flux uncertainties.
For an in-depth discussion on the background and detector uncertainty parameterizations see~\cite{Abbasi_2021}.

\section{Results\label{sec:results}}

\begin{figure}[t]
    \includegraphics[width=0.5\columnwidth]{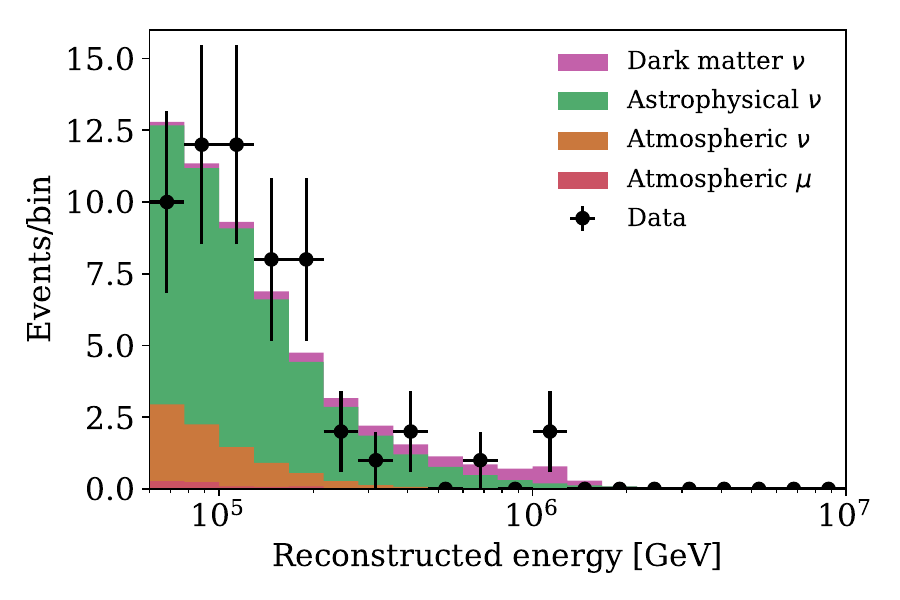}  
    \includegraphics[width=0.5\columnwidth]{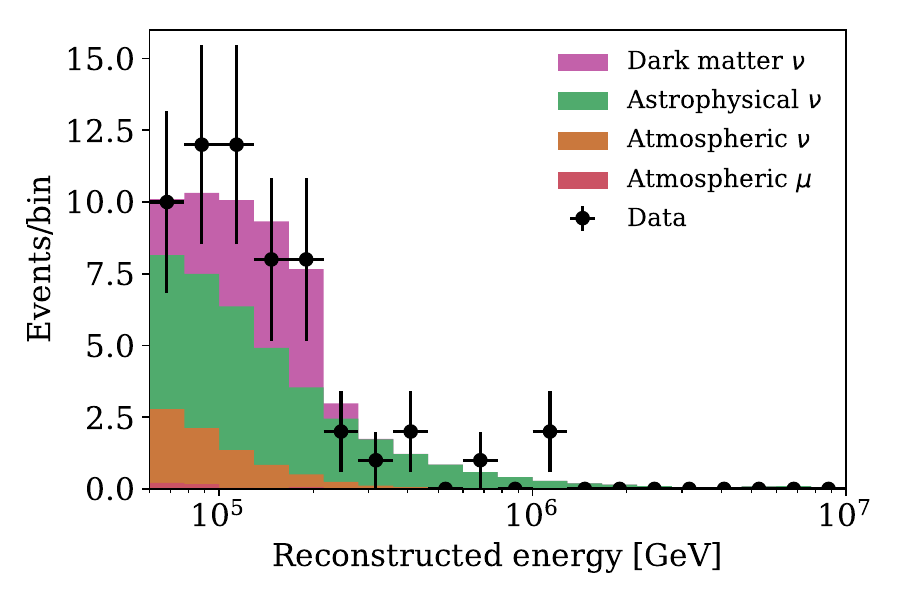}  
	\protect\caption{\textbf{\textit{Best-fit energy distributions for the annihilation and decay searches.}}
	Left panel shows the best-fit energy distribution for the annihilation scenario, which happens for $\chi\chi \to \nu\bar\nu$ channel with $m_\chi = \SI{1.23}\PeV$ and $\langle \sigma v\rangle = 2.1 \times 10^{-23}~{\rm cm}^3{\rm s}^{-1}$. An astrophysical power-law spectrum with coefficient $\gamma = 3.13$ is fitted for the annihilation case. 
	Right panel shows the best-fit energy distribution for the decay scenario, which happens for the $\chi \to b\bar b$ channel, with parameters $m_\chi = \SI{389}\TeV$ and $\tau_\chi = 2.8 \times 10^{27}{\rm s}$. An astrophysical power-law spectrum with coefficient $\gamma = 2.78$ is fitted for the decay case. This figure shows track and single cascade morphology events only. Two double cascade events were removed.}
    \label{fig:best-fit-distributions}
\end{figure}

\begin{figure}[t]
    \includegraphics[width=0.5\columnwidth]{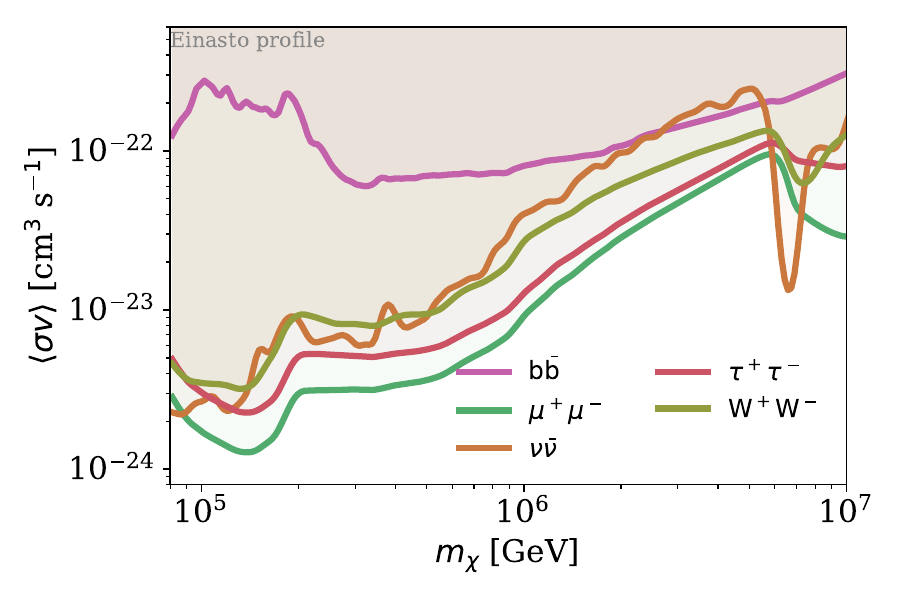}  
    \includegraphics[width=0.5\columnwidth]{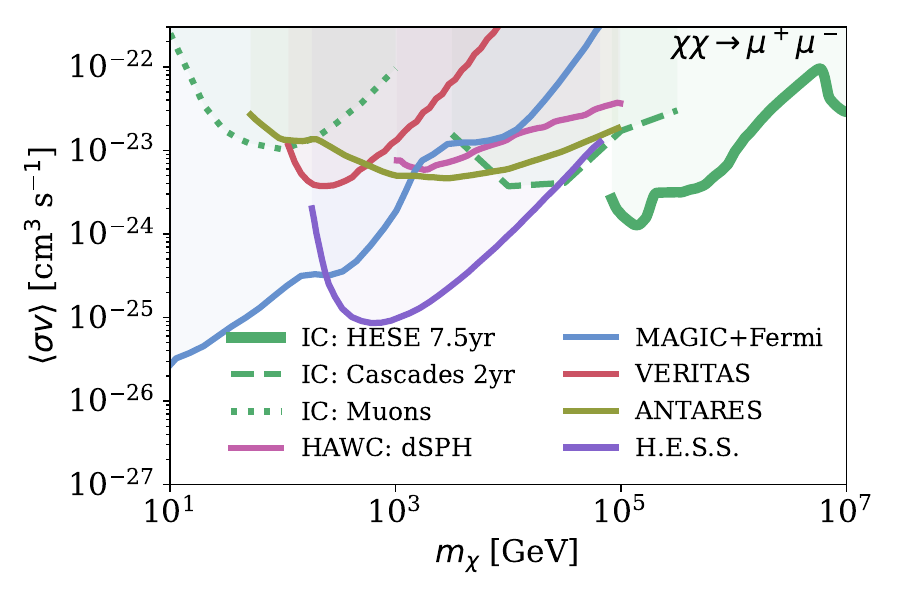}  
	\protect\caption{\textbf{\textit{Limits on the dark matter velocity-averaged self-annihilation cross-section.}}
	Left panel shows the constraints for all considered annihilation channels as a function of the dark matter mass.
	Right panel show our constraint on the $\mu^+ \mu^-$ compared with constraints from other neutrino telescopes~\cite{Albert:2016emp} and gamma-ray observatories~\cite{Aliu_2012,Ahnen:2016qkx,Abdallah:2016ygi,Albert_2018}.
	In both cases, values above the limits (i.e. shaded region) are excluded at a 90\% C.L.
	The benchmark Einasto halo profile is assumed.
	}
    \label{fig:annihilation_limits}
\end{figure}

\begin{figure}[t]
    \includegraphics[width=0.5\columnwidth]{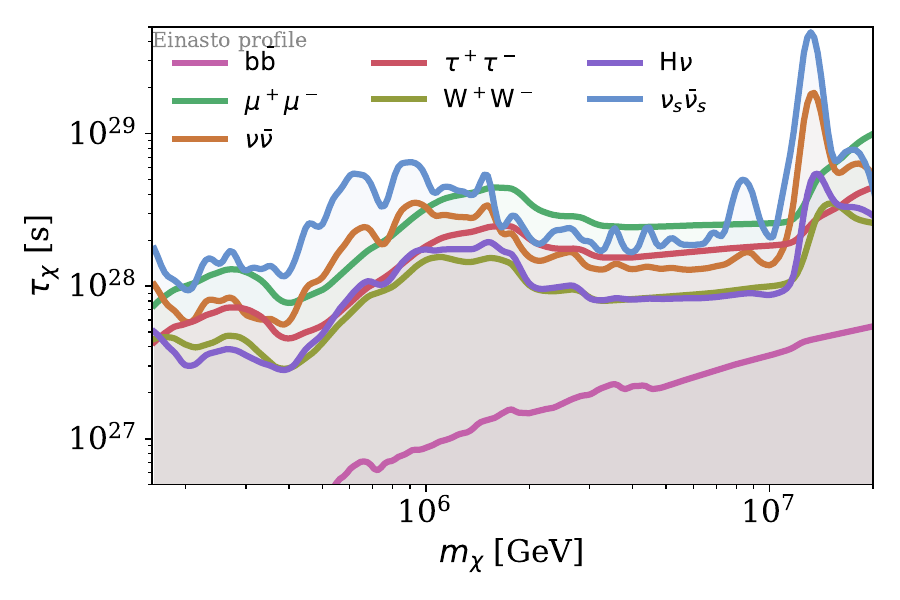}  
    \includegraphics[width=0.5\columnwidth]{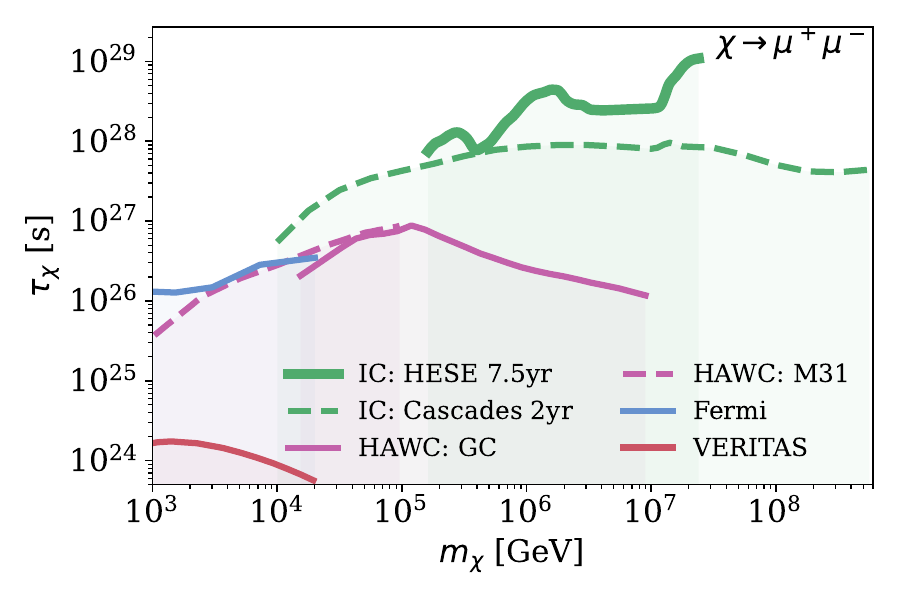}  
	\protect\caption{\textbf{\textit{Limits on the dark matter lifetime.}}
	Left panel shows the constraints for all considered decay channels as a function of the dark matter mass. Here, the $\nu_s \bar{\nu}_s$ channel represents sterile neutrinos, assuming that all steriles oscillate into a mixture of 1:1:1 SM neutrinos and do not couple to EW forces.
    The sterile neutrino spectra, therefore, has sharper features that are distinguishable from SM neutrinos.
    These limits need to be rescaled with the proper mixing angles between sterile and mass eigenstates.
	Right panel shows our constraint for the $\mu^+ \mu^-$ channel, in comparison with constraints from other neutrino telescopes~\cite{Albert:2016emp} and gamma-ray observatories~\cite{Aliu_2012,Ahnen:2016qkx,Abdallah:2016ygi,Albert_2018}.
	Lifetimes below the limits (i.e. shaded region) are excluded at a 90\% C.L. and the benchmark Einasto halo profile is assumed.
	The wiggles observed in the neutrino channels are due to the fact that this was a binned analysis and the neutrino spectra is very narrow, as shown in Fig. \ref{fig:energy_spectra}, which results in fast changes in the constraints as the signal transitions between bins.}
    \label{fig:decay_limits}
\end{figure}

None of the tested models resulted in significance greater than $2.5\sigma$ (pre-trial) when compared to the null hypothesis.
The energy distribution of the best-fit parameters for the dark matter annihilation and decay can be seen in Fig.~\ref{fig:best-fit-distributions}.
For the dark matter annihilation scenario, the best-fit point is found for the $\chi\chi \to \nu\bar\nu$ channel with $m_\chi = \SI{1.23}\PeV$ and $\langle \sigma v\rangle = 2.1 \times 10^{-23}~{\rm cm}^3{\rm s}^{-1}$. 
The long tail feature on the $\nu \bar \nu$ spectrum is due to the expected behavior from electroweak corrections \cite{Bauer_2021}.
For the dark matter decay case the best-fit point was found to be, for the $\chi \to b\bar b$ channel with parameters $m_\chi = \SI{389}\TeV$ and $\tau_\chi = 2.8 \times 10^{27}{\rm s}$.
Since these best-fit points did not result in a significant improvement over the null hypothesis we derive constraints on the dark matter contribution to the observed neutrino flux.

Despite the different treatment of the systematic uncertainties and different statistical interpretations of them, the resulting frequentist and Bayesian results were found to be numerically comparable.
To allow for better comparison with previous results, we report the frequentist results for the decay and annihilation scenarios, and the Bayesian results for the scattering scenario.

\begin{figure}[h]
    \includegraphics[width=0.5\columnwidth]{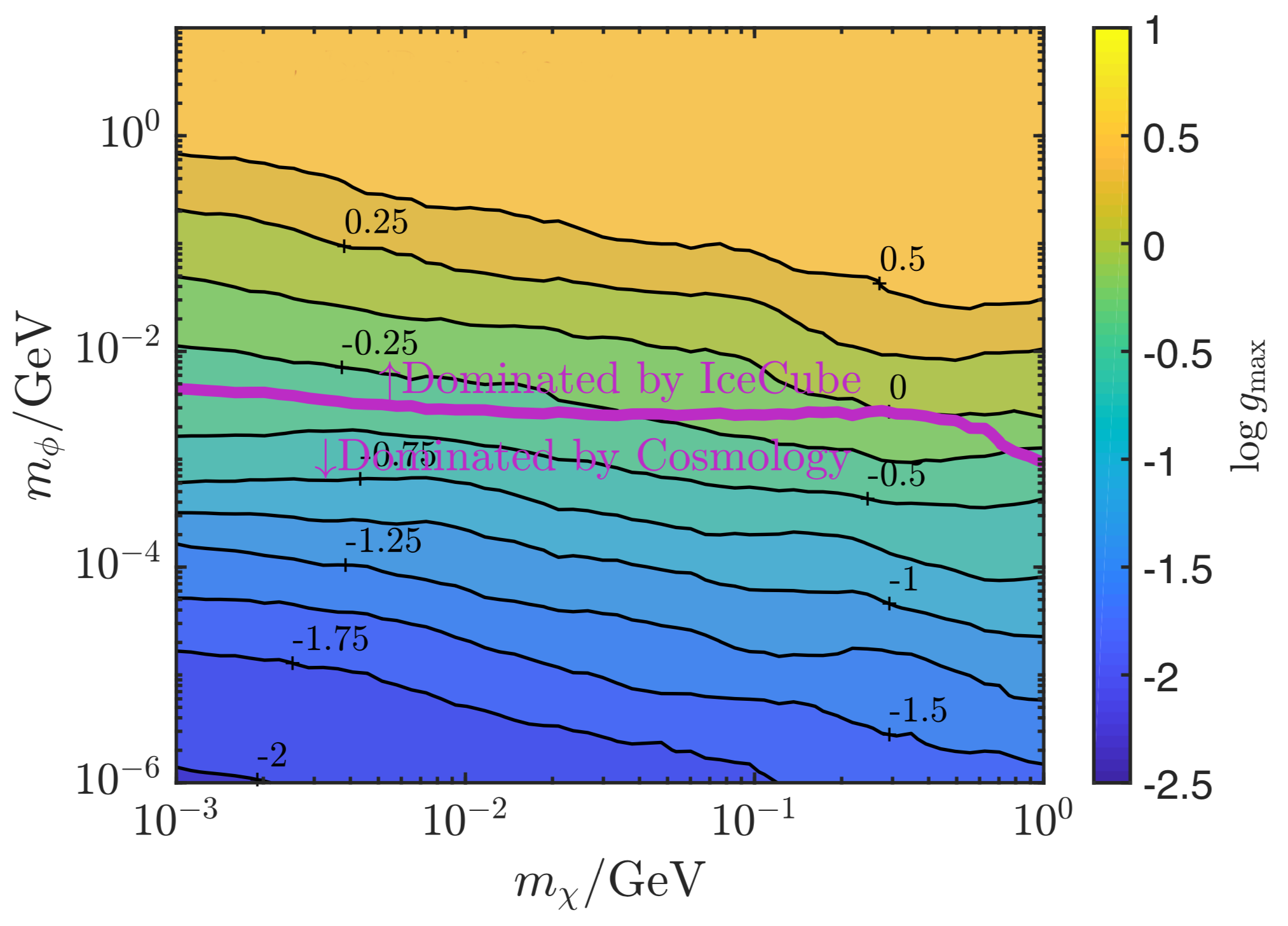}  
    \includegraphics[width=0.5\columnwidth]{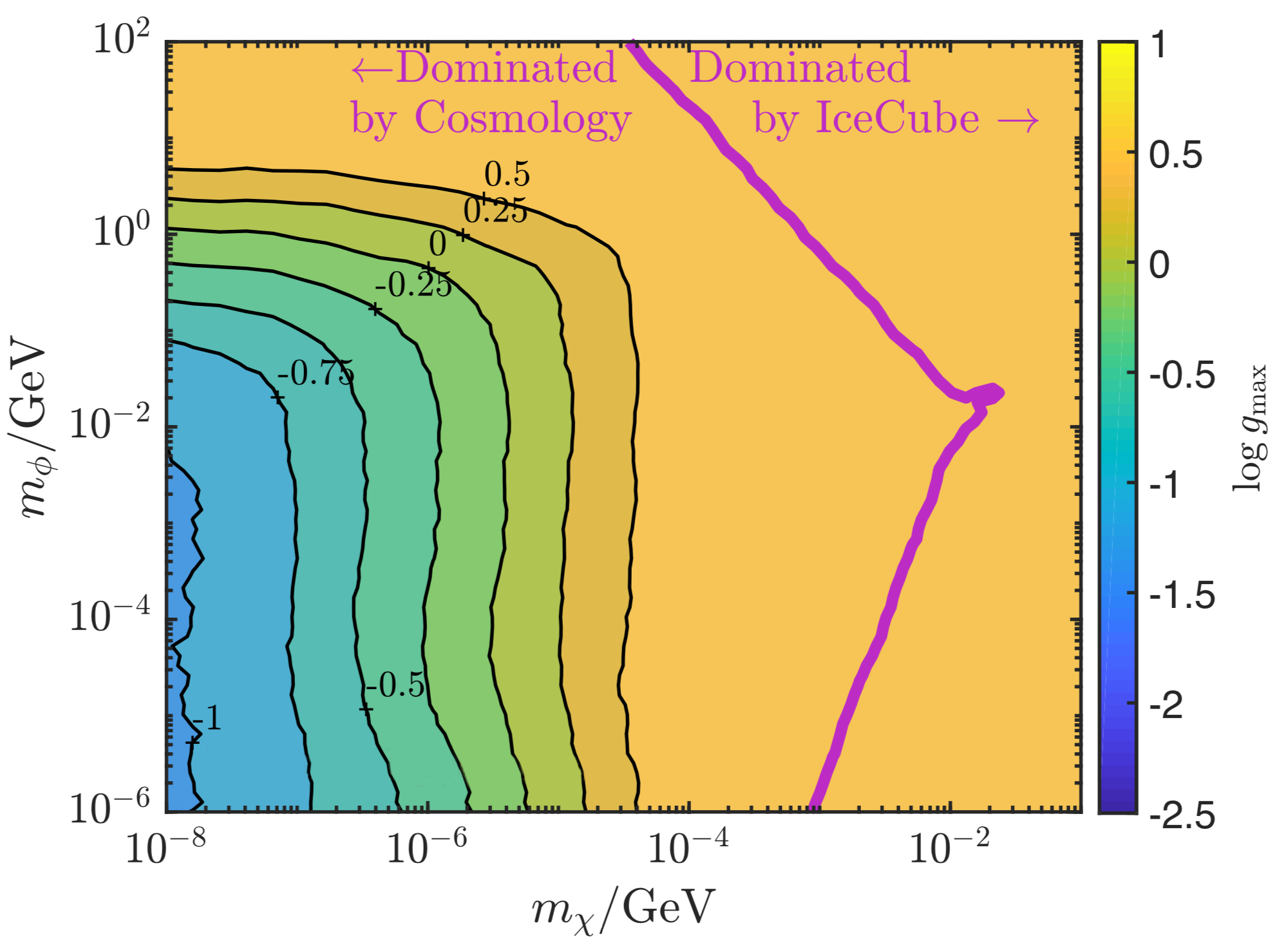}  
	\protect\caption{\textbf{\textit{Limits on the dark matter-neutrino interaction strengths for various dark matter and mediator masses for the model of a fermionic dark matter particle and a vector mediator.}}
	Left figure shows the constraints on fermionic dark matter with a vector mediator, while right shows the constraint for scalar dark matter with a fermionic mediator, in terms of the dark matter mass in the horizontal axis and the mediator mass in the vertical axis.
	The color scale shows the maximum coupling strength, $g_{\rm max} = g$ for $SF$ interaction and $g_{\rm max} = \sqrt{g_\nu g_\chi}$ for $FV$, as a 90\% credible upper limit.
	The pink line signals the region of parameter space where cosmology gives a stronger bound than the IceCube results; see for details on the cosmological constraints~\cite{Escudero:2015yka,Arguelles:2017atb}. 
	Reasons for changes relative to limits in \cite{Arguelles:2017atb} are given in the Section~\ref{sec:results}.
	The knot feature on the right panel is due to finite grid size effects.}
    \label{fig:scattering_limits}
\end{figure}

The constraints on the velocity-averaged self-annihilation cross-section for annihilation can be found in Fig.~\ref{fig:annihilation_limits} and the constraints on  dark matter lifetime for decay in Fig.~\ref{fig:decay_limits}. The constraints on dark matter-neutrino coupling constant for scattering considering the case of the fermion-vector interaction and scalar-fermion interaction in Fig.~\ref{fig:scattering_limits}. All contours are drawn as 1dof per fixed mass.

The constraints were also computed for other halo profiles: Burkert~\cite{Burkert_1995} and Navarro, Frenk \& White (NFW)~\cite{Navarro:1995iw}. 
The NFW and our baseline Einasto profiles are nearly identical and yield comparable results.
The Burkert profile yields a significantly lower dark matter density in the galactic core, implying weaker constraints.
For dark matter decay and scattering, which depend on the column density, this leads to slightly weaker constraints by around 10\%.
For dark matter annihilation, the effect is much more significant and the limits are worse by a factor as large as three.
The limits assuming the Burkert profile can be found in the Appendix~\ref{sec:appendix_profiles}.

The main source of background for this analysis are the astrophysical neutrinos.
Since the source and the spectral shape of these neutrinos are not well understood, this analysis is performed assuming a simple isotropic power-law model.
This model is found to be compatible with the observed data~\cite{Abbasi_2021}.
However, in order to test the robustness of the derived limits with respect to the astrophysical flux assumption, an isotropic double power-law astrophysical flux model is assumed.
The normalization and the power-law indices of the two components are taken as free variables in the likelihood and the limits are recomputed for a few benchmark cases.
Since additional free parameters are being introduced into the model, the limits do get weaker as expected. 
For annihilation and decay, the effect only seems to be significant for dark matter masses above $\SI{1}\PeV$.

The dark matter-neutrino results shown in Fig.~\ref{fig:scattering_limits} are qualitatively comparable to previous IceCube HESE results~\cite{Arguelles:2017atb}.
However, there is a notable distinction of the signal definition in the parameter space where cosmology gives stronger bounds than IceCube. 
This analysis includes detector systematic uncertainties and uses updated event directions \citep{Abbasi_2021}, whereas the previous one does not. 

\section{Conclusions\label{sec:conclusion}}

We have searched for indirect signs of dark matter using 7.5~years of the IceCube high-energy starting event data~\cite{Abbasi_2021}.
In our analyses, we fitted various dark matter models on top of a diffuse astrophysical neutrino flux and atmospheric backgrounds. 
An unbroken power law spectrum was assumed for the astrophysical neutrino flux. 
No evidence for dark matter was found and new stringent constraints have been obtained on the dark matter lifetime, self-annihilation cross section, and the dark matter-neutrino interaction coupling strength.

For dark matter annihilation, an upper limit on the dark matter's velocity-averaged self-annihilation cross-section was derived for various annihilation channels and for dark matter masses between $\SI{80}\TeV$ - $\SI{10}\PeV$.
Our results extend and improve existing limits, especially in the large mass region, as shown in Fig.~\ref{fig:annihilation_limits}. An improvement to the angular resolution of individual event directions contributes to the limits presented. Moreover, there is an underfluctuation of events in the direction of the galactic center compared to prior results. 

For dark matter decay, a lower limit on the dark matter lifetime has been derived for various decay channels and dark matter masses between $\SI{160}\TeV$ - $\SI{20}\PeV$.
Figure~\ref{fig:decay_limits} shows how our results improve on the previous limits derived in~\cite{Aartsen:2018mxl} and limits from other experiments in overlapping mass ranges.
Previous dark matter searches from the IceCube collaboration have largely focused on masses below $\SI{100}\TeV$ due to the model-independent upper bound from perturbative unitarity~\cite{Griest:1989wd}.
However, dark matter beyond this mass can be consistent with the observed dark matter density when a composite dark matter scenario is in play, or non-standard cosmological history is considered and/or dark matter is produced via non-thermal processes~\cite{Chung:1998zb}.
Additional dark matter searches from external collaborations using IceCube events have also explored above the 100 TeV mass range \cite{Bhattacharya:2019ucd,Chianese:2021jke,Anchordoqui:2021dls,Dekker:2019gpe,Chianese:2019kyl}.
Renewed interest in superheavy dark matter originates from the fact that collider-based searches came back empty-handed, and the superheavy dark matter region became experimentally accessible using high-energy neutrinos~\cite{Murase:2012xs,Rott:2014kfa,PhysRevD.107.103013,Kalashev_2016,PhysRevD.99.103016,B_rat_2022}.
A comprehensive discussion on cosmological implications of ultra-long lifetimes for models with larger dark matter masses is presented in \cite{PhysRevD.107.042002}.

For dark matter-neutrino scattering, an upper limit on the coupling strength has been derived for two models, one with a scalar dark matter particle and a fermionic mediator and one model with a fermionic dark matter and a vector mediator.
This model focuses on the keV-MeV mass ranges for both the mediator and the dark matter particle.
For large parts of the investigated parameter space, the limits on the coupling strengths improved upon the existing cosmological limits~\cite{Olivares-Del-Campo:2019qwe}.
Cosmological limits from dark matter-neutrino interaction consider two main effects: dark matter entropy transfer into the neutrino sector and diffusion damping
of cosmological small-scale perturbations \cite{Arguelles:2017atb}.

In this work, we have improved searches for dark matter with IceCube, but we have not yet explored the full sensitivity of the detector.
Despite not having found significant evidence for dark matter decay or annihilation, it is interesting to note that the most significant channel is the scenario where dark matter decays or annihilates into neutrinos. 
This observation motivates further searches of dark matter from the galactic center with larger data sets and improved reconstructions.
Uncertainties in the optical properties of the ice continue to limit the angular resolution of reconstructed tau and electron neutrino events at high energies.
The IceCube-Upgrade will deploy improved calibration devices which are expected to improve our understanding of the ice~\cite{Ishihara:2019aao}.
More accurate ice models will be available to re-analyze the data presented in this work once the IceCube-Upgrade is completed.
Beyond the IceCube-Upgrade, IceCube Gen-2~\cite{Aartsen_2021} will significantly increase the statistics of high-energy neutrino events and expand the reach of the analysis to higher mass dark matter. A better understanding of the backgrounds due to the diffuse astrophysical neutrino flux for example through the discovery of additional astrophysical neutrino sources and higher statistics samples obtained via IceCube-Gen2, will help discriminate diffuse extragalactic dark matter contributions. This is in particular important for the PeV range where current sensitivities show a strong dependence on the assumed astrophysical neutrino flux. Searches using extragalactic targets with large dark matter accumulations, such as clusters of galaxies will be essential to break any degeneracy between the extragalactic dark matter and astrophysical neutrino hypotheses, that cannot be resolved from the spectral shape alone.


\acknowledgments
	The IceCube collaboration acknowledges the significant contributions to this manuscript from C. Arg\"uelles, H. Dujmovic, and C. Rott. 
	USA {\textendash} U.S. National Science Foundation-Office of Polar Programs,
U.S. National Science Foundation-Physics Division,
U.S. National Science Foundation-EPSCoR,
Wisconsin Alumni Research Foundation,
Center for High Throughput Computing (CHTC) at the University of Wisconsin{\textendash}Madison,
Open Science Grid (OSG),
Extreme Science and Engineering Discovery Environment (XSEDE),
Frontera computing project at the Texas Advanced Computing Center,
U.S. Department of Energy-National Energy Research Scientific Computing Center,
Particle astrophysics research computing center at the University of Maryland,
Institute for Cyber-Enabled Research at Michigan State University,
and Astroparticle physics computational facility at Marquette University;
Belgium {\textendash} Funds for Scientific Research (FRS-FNRS and FWO),
FWO Odysseus and Big Science programmes,
and Belgian Federal Science Policy Office (Belspo);
Germany {\textendash} Bundesministerium f{\"u}r Bildung und Forschung (BMBF),
Deutsche Forschungsgemeinschaft (DFG),
Helmholtz Alliance for Astroparticle Physics (HAP),
Initiative and Networking Fund of the Helmholtz Association,
Deutsches Elektronen Synchrotron (DESY),
and High Performance Computing cluster of the RWTH Aachen;
Sweden {\textendash} Swedish Research Council,
Swedish Polar Research Secretariat,
Swedish National Infrastructure for Computing (SNIC),
and Knut and Alice Wallenberg Foundation;
Australia {\textendash} Australian Research Council;
Canada {\textendash} Natural Sciences and Engineering Research Council of Canada,
Calcul Qu{\'e}bec, Compute Ontario, Canada Foundation for Innovation, WestGrid, and Compute Canada;
Denmark {\textendash} Villum Fonden and Carlsberg Foundation;
New Zealand {\textendash} Marsden Fund;
Japan {\textendash} Japan Society for Promotion of Science (JSPS)
and Institute for Global Prominent Research (IGPR) of Chiba University;
Korea {\textendash} National Research Foundation of Korea (NRF);
Switzerland {\textendash} Swiss National Science Foundation (SNSF);
United Kingdom {\textendash} Department of Physics, University of Oxford.

\bibliographystyle{JHEP}
\bibliography{dmbiblio}
\clearpage

\appendix

\ifx \standalonesupplemental\undefined
\setcounter{page}{1}
\setcounter{figure}{0}
\setcounter{table}{0}
\setcounter{equation}{0}
\fi

\renewcommand{\thepage}{Supplemental Material-- S\arabic{page}}
\renewcommand{\figurename}{SUPPL. FIG.}
\renewcommand{\tablename}{SUPPL. TABLE}

\renewcommand{\theequation}{A\arabic{equation}}
\clearpage

\section{Comparison with other dark matter profiles\label{sec:appendix_profiles}}

\begin{figure}[h]
    \centering
    \includegraphics[width=0.5\columnwidth]{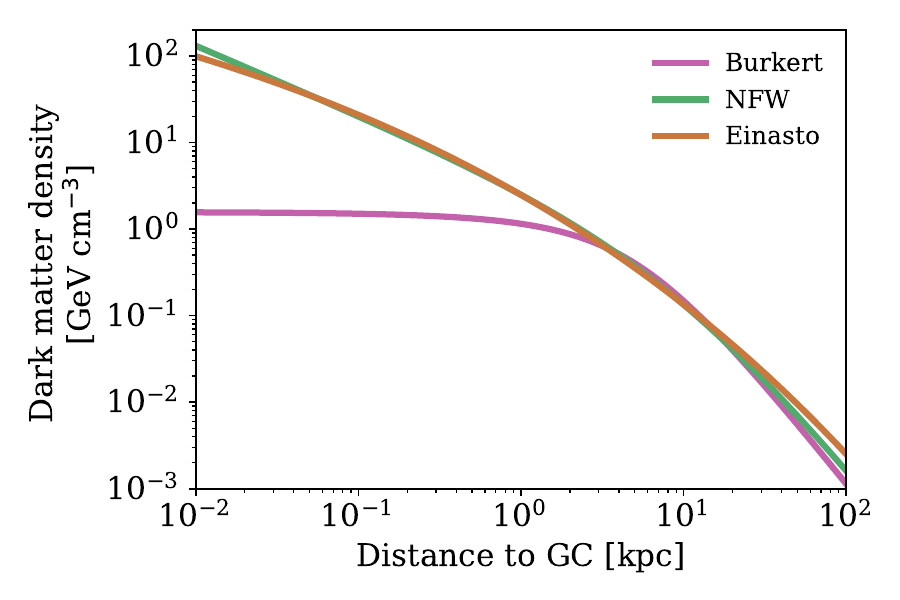}  
	\protect\caption{\textbf{\textit{Halo density profiles compared.}} The Burkert~\cite{Burkert_1995}, NFW~\cite{Navarro:1995iw}, and Einasto~\cite{einasto1965construction} profiles are shown as purple, green, and light brown colored lines.
	The Einasto and NFW profiles yield similar constraints as their profiles are alike, while the Burkert profile yields weaker constraints as its shallower around the galactic center.}
    \label{fig:profiles}
\end{figure}

\begin{figure}[ht]
    \includegraphics[width=0.5\columnwidth]{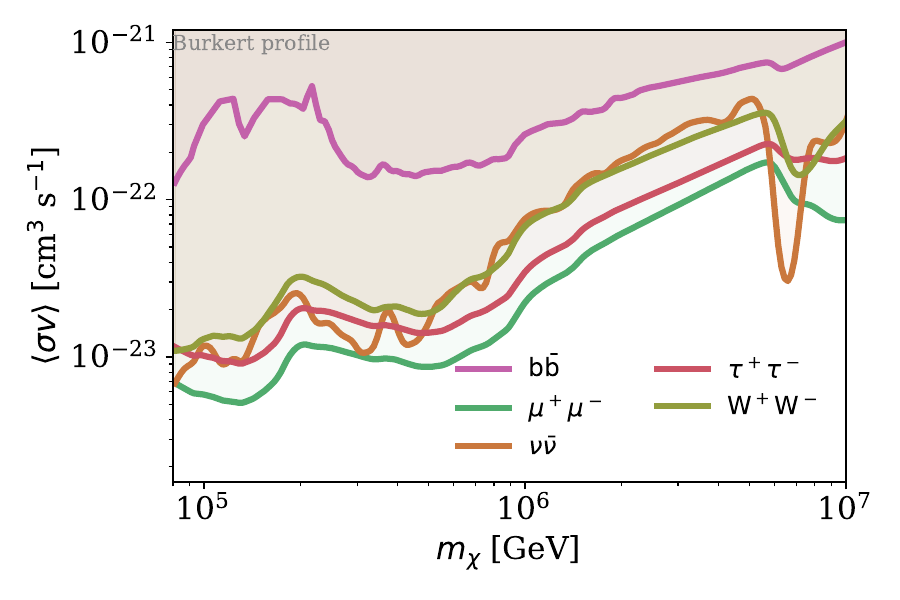}  
    \includegraphics[width=0.5\columnwidth]{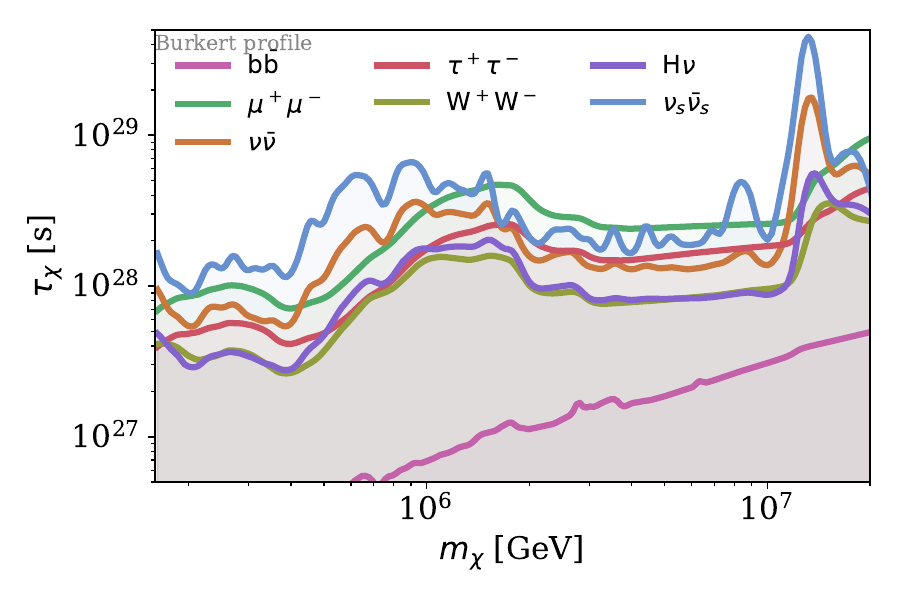}  
	\protect\caption{\textbf{\textit{Constraints on dark matter annihilation and decay using the Burkert profile.}}
	The constraints are shown for all the channels discussed in the text for the shallower profile.
	As noted in the main text the dark matter decay is minimally affected by changing the profile, but the annihilation result is weakened.
	The wiggles in the neutrino channels are statistical in nature and due to the narrow signal that migrates bins.}
    \label{fig:freq_bay_decay}
\end{figure}

The dark matter profiles studies in this work are shown in Suppl. Fig.~\ref{fig:profiles}.
The difference between the NFW~\cite{Navarro:1995iw} and Einasto~\cite{einasto1965construction} profiles is small, but the Burkert~\cite{Burkert_1995} is shallower around the galactic center.
The effect of the change in column density is only notable for dark matter annihilation, as it depends quadratically on the density.
Scattering and decay are less affected due to the only linear dependence. 
Suppl. Fig.~\ref{fig:freq_bay_decay} shows the constraints on dark matter annihilation and decay when using the Burkert profile.

\begin{table}[]
    \centering
    \begin{tabular}{c|c c c}
         Profile & Burkert~\cite{Nesti_2013} & NFW~\cite{Nesti_2013} & Einasto~\cite{Vincent_2012} \\
         \hline \vspace*{-0.4cm} \\ 
         Dark Matter Density $\rho(x)$ & $\frac{\rho_0}{(1+x)(1 + x^2)}$ & $\frac{\rho_0}{x(1+x)^2}$ & $\rho_0 \exp{(1-\frac{2x^\alpha}{\alpha})}$ \\
         Scale Radius $r_s = r / x$ & $9.26$ kpc & $16.1$ kpc & $25.7$ kpc  \\
         Local dark matter density $\rho_0(R_\odot)$ & $0.487~\text{GeV cm}^{-3}$ & $0.471~\text{GeV cm}^{-3}$ & $0.4~\text{GeV cm}^{-3}$ \\
         Distance to galactic center $R_\odot$ & $7.94$ kpc & $8.08$ kpc & $8.5$ kpc \\
         Slope parameter $\alpha$ & - & - & 0.17 \\
         
    \end{tabular}
    \caption{Parameters of dark matter profiles used in this work.} 
    \label{tab:dmprofiles}
\end{table}

\end{document}

%% file: units.tex
\DeclareSIUnit \s {\second}
\DeclareSIUnit \ns {\nano\second}
\DeclareSIUnit \mus {\micro\second}
\DeclareSIUnit \ms {\milli\second}
\DeclareSIUnit \MB {\mega\byte}
\DeclareSIUnit \GB {\giga\byte}
\DeclareSIUnit \TB {\tera\byte}
\DeclareSIUnit \PB {\peta\byte}
\DeclareSIUnit \Mbps {\mega\bit/\s}
\DeclareSIUnit \Gbps {\giga\bit/\s}
\DeclareSIUnit \Tbps {\tera\bit/\s}
\DeclareSIUnit \Pbps {\peta\bit/\s}
\DeclareSIUnit \kton {\kilo\tonne} 
\DeclareSIUnit \kt {\kilo\tonne}
\DeclareSIUnit \Mt {\mega\tonne}
\DeclareSIUnit \eV {\electronvolt}
\DeclareSIUnit \keV {\kilo\electronvolt}
\DeclareSIUnit \MeV {\mega\electronvolt}
\DeclareSIUnit \GeV {\giga\electronvolt}
\DeclareSIUnit \PeV {\peta\electronvolt}
\DeclareSIUnit \EeV {\exa\electronvolt}
\DeclareSIUnit \m {\meter}
\DeclareSIUnit \cm {\centi\meter}
\DeclareSIUnit \in {\inchcommand}
\DeclareSIUnit \km {\kilo\meter}
\DeclareSIUnit \kV {\kilo\volt}
\DeclareSIUnit \kW {\kilo\watt}
\DeclareSIUnit \MW {\mega\watt}
\DeclareSIUnit \MHz {\mega\hertz}
\DeclareSIUnit \mrad {\milli\radian}
\DeclareSIUnit \year {years}
\DeclareSIUnit \POT {POT}
\DeclareSIUnit \sig {$\sigma$}
\DeclareSIUnit\parsec{pc}
\DeclareSIUnit\lightyear{ly}
\DeclareSIUnit\foot{ft}
\DeclareSIUnit\ft{ft}
\DeclareSIUnit \ppb{ppb}
\DeclareSIUnit \ppt{ppt}
\DeclareSIUnit \samples{S}
\DeclareSIUnit \pe{PE}
\DeclareSIUnit \sr{\steradian}
\DeclareSIUnit \Mtons{Mtons}

